\documentstyle[12pt]{article}
\newcommand{\be}{\begin{equation}}
\newcommand{\ee}{\end{equation}}
\newcommand{\bea}{\begin{eqnarray}}
\newcommand{\eea}{\end{eqnarray}}
\newcommand{\ba}{\begin{array}}
\newcommand{\ea}{\end{array}}

\newcommand{\slashl}[1]{\not{\!\!#1}}

\newcommand{\norsl}{\normalsize\sl}
\newcommand{\norsc}{\normalsize\sc}
\begin{document}

\begin{titlepage}

\title{More on the Burkhardt-Cottingham Sum Rule in QCD}

\author{
\norsc  Jiro KODAIRA\thanks{Supported in part by
          the Monbusho Grant-in-Aid for Scientific Research
          No. C-05640351.} \\
\norsl  Dept. of Physics, Hiroshima University\\
\norsl  Higashi-Hiroshima 724, JAPAN\\
\\
\norsc  Satoshi MATSUDA and
         Tsuneo UEMATSU\thanks{Supported in part by
          the Monbusho Grant-in-Aid for Scientific Research
          No. C-06640392 and \quad No.06221239.} \\
\norsl  Dept. of Fundamental Sciences\\
\norsl  Faculty of Integrated Human Studies, Kyoto University\\
\norsl  Kyoto 606-01, JAPAN\\
\\
\norsc  Ken SASAKI\thanks{e-mail address:
a121004@c1.ed.ynu.ac.jp}\\
\norsl  Dept. of Physics, Yokohama National University\\
\norsl  Yokohama 240, JAPAN}

\date{}
\maketitle

\begin{abstract}
{\normalsize
\noindent
The QCD higher order effects to the polarized structure
function $g_2(x, Q^2 )$ are reanalyzed for massive quarks
in the context of the operator product expansion. We confirm
that the lowest moment of $g_2 (x, Q^2 )$ which corresponds to the
Burkhardt-Cottingham sum rule does {\sl not} suffer from radiative
corrections in perturbative QCD.}
\end{abstract}

\begin{picture}(5,2)(-300,-615)
\put(2.3,-110){KUCP-70}
\put(2.3,-125){HUPD-9410}
\put(2.3,-140){YNU-HEPTh-94-105}
\put(2.3,-155){August, 1994}
\end{picture}

\thispagestyle{empty}
\end{titlepage}
\setcounter{page}{1}
\baselineskip 24pt

In the polarized leptoproduction, we have two structure functions
$g_1$ and $g_2$.
The QCD effects on $g_1$ have been extensively studied
mainly in connection with the problem of so-called
\lq\lq spin crisis \rq\rq \,.
On the other hand, a new measurement of $g_2$
is expected to be performed at CERN, SLAC and DESY in the near
future.
Such experiment is
very important since it has been known that the
twist-3 operators as well as the twist-2 operators contribute
to $g_2$~\cite{HM} and furthermore we can check the
well-known sum rule
\be
       \int _0 ^1 dx \,\, g_2 (x, Q^2 ) = 0
\label{BC}
\ee
called the Burkhardt-Cottingham (BC) sum rule.

The fact that the twist-3 operators also contribute to the moment
of $g_2$ in the leading order of  $1/Q^2 $  produces new features
which
do not appear in the analyses of other structure functions.
At the higher-twist level, in general, the appearance
of the composite operators in the operator
product expansion (OPE) which are proportional to
the {\norsl equation of motion} makes the operator mixing problems
complicated~\cite{POLI}.
This problem has been discussed by many
authors~\cite{SVETAL,JAFFE,KOD3}
after the old papers~\cite{ARS,KOD1,KOD2}
on the polarized process.
However, as far as the first moment ($n=1$) of $g_2$ is concerned,
the above complexity is irrelevant since there is no operators
corresponding to $n=1$. Therefore it is naively expected that the
BC sum rule is exact at least perturbatively.

Recently two groups~\cite{MN,ALNR} have discussed the QCD effects
at order $\alpha _{\rm s}$ to BC sum rule for massive
quarks and reached different
conclusions on the validity of BC sum rule Eq.(\ref{BC}).
This controversial situation must be resolved especially for the
experimentalists who plan the measurement of $g_2$. The purpose
of this note is to present an independent calculation of
the QCD corrections to BC sum rule in the framework of OPE and
try to settle the above issue.

Spin-dependent structure functions are defined by the
antisymmetric (A) part of the Fourier
transform of the commutator of two electromagnetic currents
sandwiched between polarized nucleon states.
\[
  W_{\mu\nu} = \frac{1}{2\pi} \int d^4 x e^{iq\cdot x}
     \left\langle p,s | [J_{\mu}(x) , J_{\nu}(0)]|p,s \right\rangle
          \equiv W^S_{\mu\nu} + i W^A_{\mu\nu} ,
\]
where $p$ ($s$) is the nucleon's momentum (covariant spin) and $q$
is the
virtual photon momentum. We also introduce the current correlation
function
\be
  T_{\mu\nu} = i \int d^4 x e^{iq\cdot x} \left\langle p,s |
       T(J_{\mu}(x) J_{\nu}(0))|p,s \right\rangle \equiv
                       T^S_{\mu\nu} + i T^A_{\mu\nu} ,
\label{T}
\ee
such that
\[
            W_{\mu\nu} = \frac{1}{\pi}{\rm Im}T_{\mu\nu} .
\]
The antisymmetric part $W^A_{\mu\nu}$ is expressed in terms of
two structure functions $g_1$ and $g_2$.
\[
  W^A_{\mu\nu} =  \varepsilon _{\mu\nu\lambda\sigma} q^{\lambda}
                \left\{ s^{\sigma} \frac{1}{p\cdot q} g_1(x, Q^2 )
                + ( p\cdot q s^{\sigma} - q\cdot s p^{\sigma} )
               \frac{1}{(p\cdot q)^2} g_2 (x, Q^2 )\right\},
\]
where $x$ is the Bjorken variable $x=Q^2 /2 p\cdot q$ and $q^2 = - Q^2$.
We will make the same tensor decomposition for also the current
correlation function.
\be
  T^A_{\mu\nu} =  \varepsilon _{\mu\nu\lambda\sigma} q^{\lambda}
                \left\{ s^{\sigma} \frac{1}{p\cdot q} t_1(\omega, Q^2 )
                + ( p\cdot q s^{\sigma} - q\cdot s p^{\sigma} )
                \frac{1}{(p\cdot q)^2} t_2 (\omega, Q^2 )\right\},
\label{ca}
\ee
where $\omega = 1/x$.

According to OPE, the current correlation function Eq.(\ref{T}) is
written as follows in the Bjorken limit~\cite{KOD3},
\bea
 T^A_{\mu\nu} &=& - \varepsilon _{\mu\nu\lambda\sigma} q^{\lambda}
              \sum_{n:{\rm odd}} \left( \frac{2}{Q^2 }\right )^n
                     q_{\mu _1} \cdots q_{\mu _{n-1}}\nonumber\\
     & & \qquad \times
    \Bigl\{ E_q^n \langle p,s | R_q^{\sigma\mu_{1}\cdots \mu_{n-1}}
              |p,s \rangle
    + \sum_j E_j^n \langle p,s | R_j^{\sigma\mu_{1}\cdots \mu_{n-1}}
         |p,s \rangle \Bigr\} .
\label{ope}
\eea
$R_i$'s are the composite operators and $E_i$'s the corresponding
coefficient
functions. In Eq.(\ref{ope}), $R_q$ are the twist 2 operator
and others the twist 3 ones. For simplicity, let us consider
the flavor non-singlet case.
$R_q$ are explicitly given by the following traceless operators.
(Subtractions of trace terms are always understood in the
following.)
\[
  R_q^{\sigma\mu_{1}\cdots \mu_{n-1}} =
         i^{n-1} \overline{\psi}\gamma_5 \gamma^{\{\sigma}D^{\mu_1}
              \cdots D^{\mu_{n-1}\}}\psi \,,
\]
where $\{ \quad \}$ denotes the symmetrization over the Lorentz
indices between them and $D^{\mu}$ is the covariant derivative.
(The flavor matrices $\lambda_i$ for the quark field $\psi$ are
suppressed in this paper.)
We have for the twist 3 operators,
\bea
  R_F^{\sigma\mu_{1}\cdots \mu_{n-1}} &=&
         \frac{i^{n-1}}{n} \Bigl[ (n-1) \overline{\psi}\gamma_5
       \gamma^{\sigma}D^{\{\mu_1} \cdots D^{\mu_{n-1}\}}\psi \nonumber\\
   & & \qquad\qquad - \sum_{l=1}^{n-1} \overline{\psi} \gamma_5
       \gamma^{\mu_l }D^{\{\sigma} D^{\mu_1} \cdots D^{\mu_{l-1}}
            D^{\mu_{l+1}} \cdots D^{\mu_{n-1}\}}
                             \psi \Bigr] \label{quark}\\
  R_m^{\sigma\mu_{1}\cdots \mu_{n-1}} &=&
          i^{n-2} m \overline{\psi}\gamma_5
       \gamma^{\sigma}D^{\{\mu_1} \cdots D^{\mu_{n-2}}
        \gamma ^{\mu_{n-1}\}} \psi \label{mass} \\
  R_k^{\sigma\mu_{1}\cdots \mu_{n-1}} &=& \frac{1}{2n}
              \left( V_k - V_{n-1-k} + U_k + U_{n-1-k} \right)
                                                           \label{gluon}
\eea
$m$ in Eq.(\ref{mass}) is the quark mass (matrix).
$V$ and $U$ in Eq.(\ref{gluon}) depend on
the gluon field strength $G_{\mu\nu}$ and the dual tensor
$\widetilde{G}^{\mu \nu}=
{1\over 2}\varepsilon_{\mu\nu\alpha\beta}G^{\alpha\beta}$, 
respectively, and given by
\bea
    V_k &=& i^n g S \overline{\psi}\gamma_5
       D^{\mu_1} \cdots G^{\sigma \mu_k } \cdots D^{\mu_{n-2}}
        \gamma ^{\mu_{n-1}} \psi \nonumber \\
    U_k &=& i^{n-3} g S \overline{\psi}
       D^{\mu_1} \cdots \widetilde{G}^{\sigma \mu_k } \cdots 
             D^{\mu_{n-2}} \gamma ^{\mu_{n-1}} \psi \nonumber
\eea
where $S$ means the symmetrization over $\mu_i$ and $g$ is the
strong
coupling constant.
The operators Eqs.(\ref{quark} - \ref{gluon}) are not all
independent
of each other but related through the operators which are
proportional
to the {\norsl equation of motion}, 
\bea
     R_{eq}^{\sigma\mu_{1}\cdots \mu_{n-1}} &=&
           i^{n-2} \frac{n-1}{2n} S [ \overline{\psi} \gamma_5
          \gamma^{\sigma} D^{\mu_1} \cdots D^{\mu_{n-2}}
        \gamma ^{\mu_{n-1}} (i\slashl{D} - m )\psi \nonumber\\
    & & \qquad\qquad\qquad\qquad + \overline{\psi} (i\slashl{D} - m )
              \gamma_5 \gamma^{\sigma} D^{\mu_1} \cdots D^{\mu_{n-2}}
        \gamma ^{\mu_{n-1}} \psi ] \nonumber \,.
\eea
The following relation is easily obtained :
\be
     R_F^{\sigma\mu_{1}\cdots \mu_{n-1}} = 
        \frac{n-1}{n} R_m^{\sigma\mu_{1}\cdots \mu_{n-1}}
             + \sum_{k=1}^{n-2} (n-1-k)
                 R_k^{\sigma\mu_{1}\cdots \mu_{n-1}} +
             R_{eq}^{\sigma\mu_{1}\cdots \mu_{n-1}} .
\label{oprelation}
\ee

Now define the matrix elements of these operators between nucleon
states
with momentum $p$ and spin $s$ by
\bea
  \langle p,s | R_q^{\sigma\mu_{1}\cdots \mu_{n-1}} |p,s \rangle
      &=& - a_n s^{\{\sigma}p^{\mu_1} \cdots p^{\mu_{n-1}\}}
                                         \label{element1}\\
  \langle p,s | R_F^{\sigma\mu_{1}\cdots \mu_{n-1}} |p,s \rangle
      &=& - \, \frac{n-1}{n}\, d_n ( s^{\sigma}p^{\mu_1}
                    - s^{\mu_1}p^{\sigma})
                    p^{\mu_2} \cdots p^{\mu_{n-1}} \\
  \langle p,s | R_m^{\sigma\mu_{1}\cdots \mu_{n-1}} |p,s \rangle
      &=& - e_n ( s^{\sigma}p^{\mu_1} - s^{\mu_1}p^{\sigma})
                    p^{\mu_2} \cdots p^{\mu_{n-1}} \\
  \langle p,s | R_k^{\sigma\mu_{1}\cdots \mu_{n-1}} |p,s \rangle
      &=& - f_n^k ( s^{\sigma}p^{\mu_1} - s^{\mu_1}p^{\sigma})
                    p^{\mu_2} \cdots p^{\mu_{n-1}} \\
  \langle p,s | R_{eq}^{\sigma\mu_{1}\cdots \mu_{n-1}} |p,s \rangle 
      &=& 0 \label{elementeq}.
\eea
We have normalized operators such that for a free
quark target $a_n = d_n = e_n = 1$. On the other hand, $f_n^k
= {\cal O} (g^2 )$.
The moment sum rule for $g_1$ and $g_2$ become
using Eq.(\ref{element1} - \ref{elementeq}),
\be
      \int_0^1 dx x^{n-1} g_1 (x,Q^2 ) = \frac{1}{2}
                          a_n E_q^n (Q^2 ).
\label{g1sumrule}
\ee
\bea
      \int_0^1 dx x^{n-1} g_2 (x,Q^2 ) &=& - \frac{n-1}{2n}
           a_n E_q^n (Q^2 )  \nonumber\\
           & &  + \frac{1}{2} \left[ 
                    \frac{n-1}{n}d_n E_F^n (Q^2)+e_n E_m^n (Q^2 )
                    + \sum_k f_n^k E_k^n (Q^2 ) \right] .
\label{g2sumrule}
\eea
It is to be noted that from Eq.(\ref{oprelation}) we have
the following constraint,
\[ \frac{n-1}{n} d_n = \frac{n-1}{n} e_n + \sum_{k=1}^{n-2}
               (n-1-k) f_n^k .\]

As mentioned before, we see from Eqs.(\ref{quark}-\ref{gluon}) that
twist 3 operators can {\sl not} be defined for $n=1$. 
Namely, $d_n$, $e_n$, $f_n^k$'s are identically zero for $n=1$ in 
(\ref{g2sumrule}).
Also we have
a \lq\lq kinematical\rq\rq\ factor $n-1$ in front of the
contribution
from the twist 2 operators in the moment sum rule for $g_2$
Eq.(\ref{g2sumrule}). Therefore the OPE analysis implies that
the BC sum rule does not suffer from the radiative corrections at
all.

Here note that when expanding $t_1(\omega ,Q^2)$ and $t_2(\omega ,Q^2)$
in Eq.(\ref{ca}) in powers of $\omega$, their n-th coefficients
are just equal to the moment sum rules Eq.(\ref{g1sumrule}) and
Eq.(\ref{g2sumrule}) respectively.
Therefore if we know $t_2(\omega ,Q^2)$,
we can check whether the above {\it na\"{\i}ve} argument for $n=1$
holds or not. (This is the standard procedure to obtain
the higher order corrections for the coefficient functions.)
The order of $\alpha_s$ diagrams contributing to $T^A_{\mu\nu}$
are shown in Fig.1. The calculation is performed with massive
quarks. The ultraviolet divergences are regularized with momentum
cut
off $\Lambda$. To regularize the infrared singularities, we give a
mass $\lambda$ to the gluon. Note that the collinear singularities
are already regulated by quark mass. We choose the on-shell
renormalization scheme.
Although we have calculated the diagrams without making
any approximations for $m^2 / Q^2$, we present only the results
after taking the limit $m^2 / Q^2 \rightarrow 0$ in accordance with
the fact that we kept only the leading twist terms in the OPE.
The contributions from each diagrams contain current non-conserved
terms which cancel out after adding all contributions.
So we will neglect such terms in the following results.

The results for each diagrams read ($C_F(=4/3)$ is the Casimir
operator for quarks):

\noindent
(a) Born + self-energy contribution of Fig.1a :
\bea
    t_1^{({\rm a})} &=& \sum_n \omega ^n \biggl[ 1 +
\frac{g^2}{16\pi^2}
         C_F \biggl\{ \ln \frac{Q^2}{m^2} + 3 + \frac{1}{n}
           - \sum_{r=1}^n \frac{1}{r} + 2 \ln
                  \frac{\lambda^2}{m^2} \biggr\}\biggr]\,,\nonumber
\\
    t_2^{({\rm a})} &=& 0 \,.\nonumber
\eea
(b) vertex contribution of Fig.1b :
\bea
    t_1^{({\rm b})} &=& \frac{g^2}{8\pi^2} C_F \sum_n \omega ^n
               \biggl[ \left( -1 - 2 \sum_{r=2}^n \frac{1}{r} \right)
             \ln \frac{Q^2}{m^2} -4 + \frac{1}{n} + 2
                   \sum_{r=1}^n \frac{1}{r} \nonumber\\
       & & \qquad\qquad\qquad\qquad\qquad - 2 \sum_{r=1}^n
          \frac{1}{r^2} - 2 \sum_{s=1}^n \frac{1}{s}\sum_{r=1}^s
           \frac{1}{r} -2 \ln \frac{\lambda^2}{m^2} \biggr] \,,\nonumber\\
    t_2^{({\rm b})} &=& \frac{g^2}{8\pi^2} C_F \sum_n \omega ^n
               \biggl[ \left( -1 + \frac{1}{n} \right)
             \ln \frac{Q^2}{m^2} + \frac{1}{2} - \frac{3}{n}
              + \frac{1}{n^2}
        + \left( \frac{1}{2} + \frac{1}{n} \right)
                    \sum_{r=1}^n \frac{1}{r} \biggr]\,.\nonumber
\eea
(c)box contribution of Fig.1c :
\bea
    t_1^{({\rm c})} &=& \frac{g^2}{8\pi^2} C_F \sum_n \omega ^n
               \biggl[ \frac{1}{n(n+1)}
             \ln \frac{Q^2}{m^2} - \frac{4}{n} + \frac{4}{n+1}
              + \frac{1}{n^2} \nonumber\\
       & & \qquad\qquad\qquad\qquad - \frac{2}{(n+1)^2} + \left( 2 +
               \frac{1}{n(n+1)} \right)
        \sum_{r=1}^n \frac{1}{r} + \ln \frac{\lambda^2}{m^2} \biggr]
                   \,,\nonumber\\
    t_2^{({\rm c})} &=& \frac{g^2}{8\pi^2} C_F \sum_n \omega ^n
               \biggl[ \left( \frac{2}{n+1} - \frac{1}{n} \right)
             \ln \frac{Q^2}{m^2} + \frac{3}{n} - \frac{6}{n+1}
              + \frac{1}{n^2}\nonumber\\
   & & \qquad\qquad\qquad\qquad + \frac{4}{(n+1)^2}
                    + \left( \frac{2}{n+1}
                    - \frac{1}{n} \right)
                    \sum_{r=1}^n \frac{1}{r} \biggr]\nonumber\,.
\eea
Now the entire expressions for $t_1$ and $t_2$ 
in which the above three contributions are added together are:
\bea
  t_1 &=& \sum_n \omega ^n \biggl[ 1 + \frac{g^2}{8\pi^2} C_F
        \biggl\{ - \frac{1}{2} \left( 1- \frac{2}{n(n+1)}
         + 4 \sum_{r=2}^n \frac{1}{r} \right) \ln \frac{Q^2}{m^2}
          \nonumber\\
      & & \qquad\qquad - \frac{5}{2} - \frac{5}{2n} + \frac{4}{n+1}
            + \frac{1}{n^2} - \frac{2}{(n+1)^2}
                                \label{g1result}\nonumber \\
      & & \qquad\qquad + \left( \frac{7}{2} + \frac{1}{n(n+1)} \right)
        \sum_{r=1}^n \frac{1}{r} - 2 \sum_{r=1}^n \frac{1}{r^2}
            - 2 \sum_{r=s}^n \frac{1}{s}\sum_{r=1}^s \frac{1}{r}
                   \biggr\} \biggr] \nonumber \,,\\
  t_2 &=& \frac{g^2}{8\pi^2} C_F \sum_n \omega ^n \biggl[
         - \frac{1}{2}\,\, \frac{2(n-1)}{n+1} \ln \frac{Q^2}{m^2}\nonumber\\
      & & \qquad\qquad\qquad\qquad + \frac{1}{2} - \frac{6}{n+1}
                + \frac{4}{(n+1)^2} 
       + \frac{n+5}{2(n+1)} \sum_{r=1}^n \frac{1}{r}
                      \biggr] \label{g2result} .
\eea
Eq.(\ref{g1result}) and its interpretation is in agreement with the
result of Ref.~\cite{MN}. However
The result Eq.(\ref{g2result}) is in agreement with that of
Ref.~\cite{ALNR}. The first moment $(n=1)$ which corresponds to the BC
sum rule vanishes. So our calculation reconfirms the validity of
the BC sum rule and shows that the OPE analysis is consistent with
the QCD perturbation theory.

In conclusion we have calculated the virtual Compton scattering
amplitude
at order $\alpha_{\rm s}$ and shown that the BC sum rule does not
receive
any corrections in perturbative QCD based on the OPE. 
In this respect, here we note that the BC sum rule is not 
only protected from QCD radiative corrections 
but also free from target mass effects \cite{MU}.

Finally, we expect that future experiments on $g_2$ will confirm
the BC sum rule in its original form.

\vspace{2.5cm}
We would like to thank Yoshiaki Yasui for discussions.


\newpage
\noindent
{\large Figure Caption}
\baselineskip 16pt

\vspace{1.5cm}
\noindent
Fig.1

\noindent
Diagrams (a) self-energy, (b) vertex and (c) box, contributing to
the current correlation function at order $g^2$.

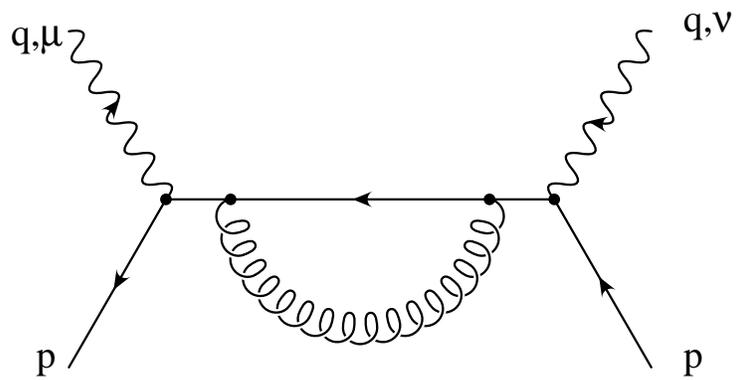
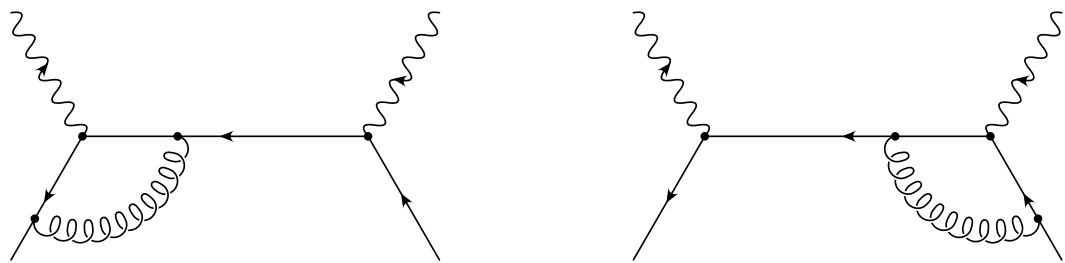
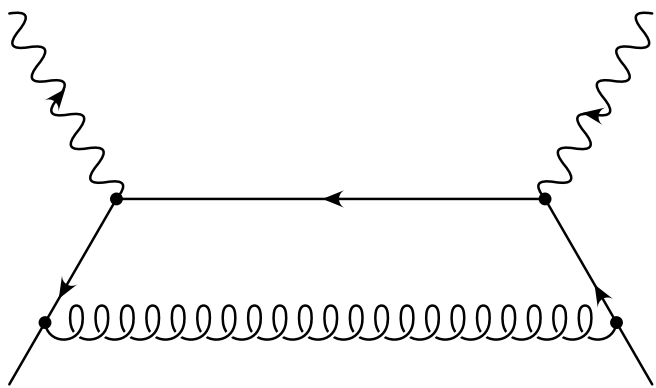
\end{document}